\documentclass[preprint,amsmath,amssymb,aps]{revtex4-2}

\usepackage{graphicx}
\usepackage{bm}
\usepackage{url}
\usepackage{hyperref}
\usepackage{color,soul} 

\DeclareMathOperator{\Tr}{Tr}

\begin{document}
\title{Frank-Read Mechanism in Nematic Liquid Crystals}

\author{Cheng Long$^{1,2}$}
\author{Matthew J. Deutsch$^{1}$}
\author{Joseph Angelo$^{1}$}
\author{Christopher Culbreath$^{1}$}
\author{Hiroshi Yokoyama$^{1,2}$}\email{hyokoyam@kent.edu (co-corresponding author)}
\author{Jonathan V. Selinger$^{1,2}$}\email{jselinge@kent.edu (co-corresponding author)}
\author{Robin L. B. Selinger$^{1,2}$}\email{rselinge@kent.edu (co-corresponding author)}

\affiliation{1. Advanced Materials and Liquid Crystal Institute, Kent State University, Kent, Ohio 44242, USA \\ 2. Physics Department, Kent State University, Kent, Ohio 44242}

\date{\today}

\begin{abstract} 
In a crystalline solid under mechanical stress, a Frank-Read source is a pinned dislocation segment that repeatedly bows and detaches, generating concentric dislocation loops. We demonstrate that in nematic liquid crystals, an analogous Frank-Read mechanism can generate concentric disclination loops. Using experiment, simulation, and theory, we study a disclination segment pinned between surface defects on one substrate in a nematic cell. Under applied twist of the nematic director, the pinned segment bows and emits a new disclination loop which expands, leaving the original segment intact; loop emission repeats for each additional 180$^\circ$ of applied twist. We present experimental micrographs showing loop expansion and snap-off, numerical simulations of loop emission under both quasistatic and dynamic loading, and theoretical analysis considering both free energy minimization and the balance of competing forces. We find that the critical stress for disclination loop emission scales as the inverse of segment length, and changes as a function of strain rate and temperature, in close analogy to the Frank-Read source mechanism in crystals. Lastly, we discuss how Frank-Read sources could be used to modify microstructural evolution in both passive and active nematics.

\end{abstract}

\maketitle

\section{Introduction}
In both crystalline solids and nematic liquid crystals, the nucleation, motion, annihilation and microstructure of line defects influence macroscopic material properties. In crystals, dislocations disrupt translational order; ~\cite{Weertman1964,Anderson2017}  under external stress, their motion produces plastic yield. In nematic liquid crystals, disclinations disrupt orientational order~\cite{Kleman1983,Kleman1989,Kleman2008,Alexander2012,Tang2017,Tang2019,Long2021}.  Disclinations in nematics nucleate, move, and annihilate in response to elastic stress~\cite{Long2021} and fluid flow~\cite{Mather1996,Copar2021}.
In passive liquid crystals, disclinations form during cooling into the nematic phase, or when surface anchoring is broken \cite{y1_angelo_2017}, and may be stabilized by topological constraints such as those associated with colloidal inclusions \cite{y2_copar_2015} and patterned surface anchoring ~\cite{y3_fleury_2009,y4_yokoyama_2010,y5_wang_2017}. In active nematics, disclinations spontaneously nucleate, move, and annihilate ~\cite{Marchetti2013,Giomi2013,Giomi2014,Kumar2018,Shankar2019,Duclos2020,Binysh2020}.  Mechanisms governing disclination nucleation are therefore important to understand microstructural evolution in nematics.

In crystalline solids, the most common mechanism for generating dislocations is a Frank-Read source~\cite{Weertman1964,Anderson2017,Frank1950}.  Here, a pre-existing dislocation line is pinned at two points along its length, e.g. by impurities or entanglement with other defects.  Under applied stress, the dislocation segment bows out into a circular arc.  As stress increases, the arc expands and eventually self-intersects; a new dislocation loop snaps off, while the segment between the pinning points remains intact.  This process can repeat periodically, leading to formation of a series of expanding concentric dislocation loops, resulting in plastic deformation of the crystal under stress. This continuous process may arrest if outgoing loops are blocked, creating a back stress on the source.

\begin{figure}[b]
\includegraphics[width=\columnwidth]{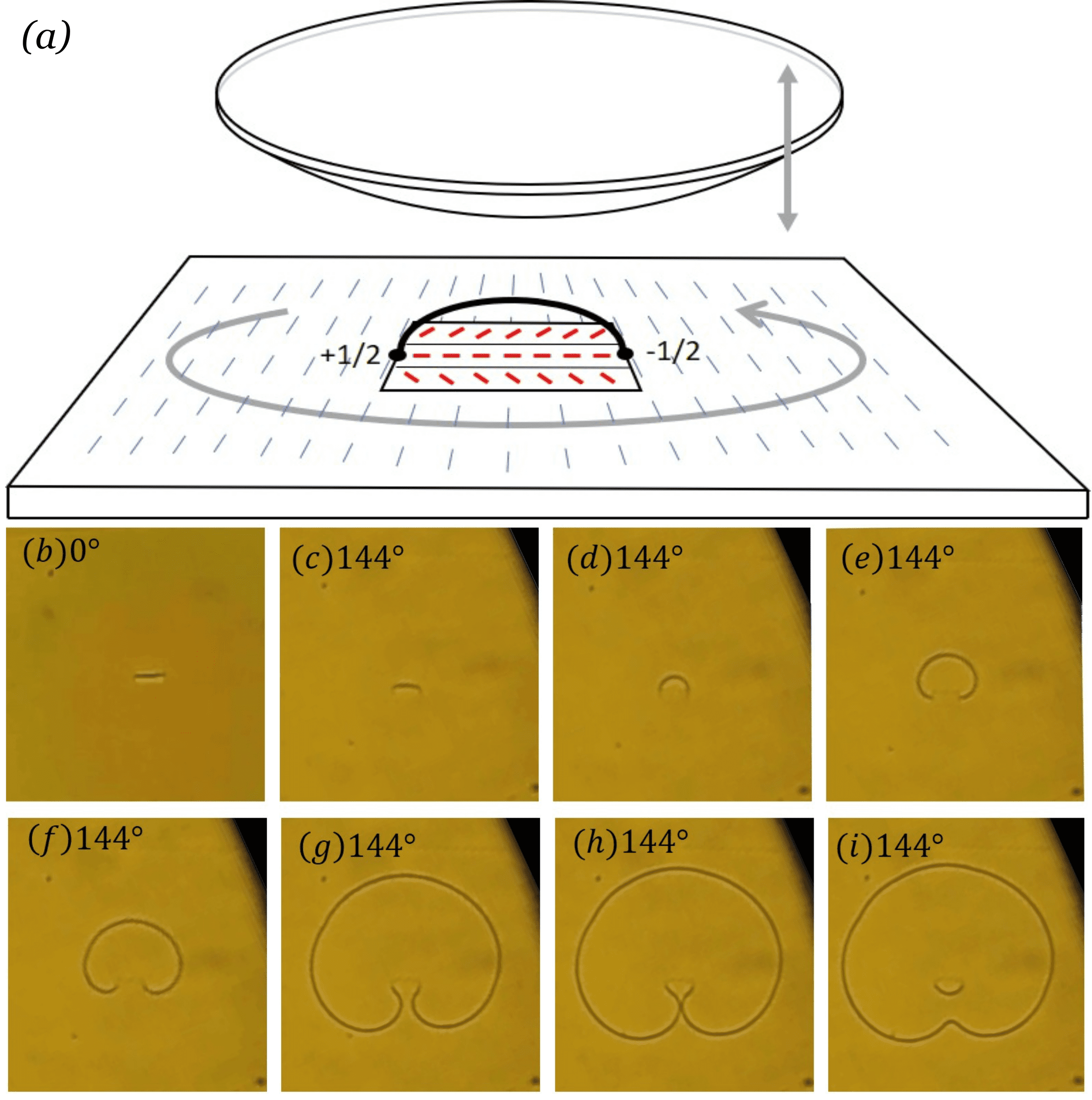}
\caption{Experimental micrographs showing a Frank-Read source in a dynamic liquid crystal cell, with a disclination segment pinned at two point defects on the lower substrate, before and after an imposed twist strain. (a) Diagram of the dynamic cell. The surface of a convex lens (upper substrate) imposes uniform planar alignment. The anchoring pattern with two point defects is shown on the lower substrate. (b) Viewed from above, the arch appears as a straight line at $t = 0\, s$. (c)--(i) Disclination shape evolution after the angle between the two substrates is increased via mechanical rotation by $144^{\circ}$, showing the Frank-Read mechanism. The defect segment initially bows outward, shown at times (c) $t=210\,s$ (d) $t=300\,s$ (e) $t=400\,s$ (f) $t=540\,s$ (g) $t=740\,s$. (h) Self-intersection occurs at $t=770\,s$. (i) At  $t=788\,s$, a new loop has snapped off, leaving the initial segment intact.}
\label{fig:experiment}
\end{figure}

Here we explore an analogous Frank-Read mechanism arising in nematic liquid crystals. In crystals, the Frank-Read mechanism has been extensively investigated to understand material susceptibility to plastic deformation~\cite{Anderson2017}.
To gain insight into the Frank-Read mechanism in nematics, we perform experiments on a pinned disclination segment under twist deformation, and perform simulations of disclination loop emission under both quasistatic and dynamic strain conditions. Next we develop an analytic theory using minimization of a two-dimensional (2D) free energy, and a more intuitive theory using the Peach-Koehler force, which is the force driving defect motion due to local stress.  Lastly, we discuss the potential role of engineered Frank-Read sources to drive heterogeneous loop nucleation and control microstructural evolution in passive and active nematics. 

\section{Experiment}

To study a Frank-Read source in a nematic, we perform an experiment using a dynamic cell~\cite{y6_culbreath_2015, y7_angelo_2017} where the in-plane rotation angle between top and bottom glass substrates can be changed to impose twist. The upper substrate, which is a convex lens with radius of curvature of two meters, has uniform planar anchoring, with a polyimide layer aligned via uniaxial rubbing. The lower substrate has an anchoring layer with two point defects of topological charge $+1/2$ and $-1/2$ ~\cite{ Kleman1983,y4_yokoyama_2010,y7_angelo_2017}, as shown in Fig.~\ref{fig:experiment} (a), patterned via a photoalignment technique \cite{y8_culbreath_2011}. The region with defects on the lower substrate consists of three areas with distinct orientations, surrounded by uniform planar anchoring, producing the two point defects \cite{Kleman1983, y4_yokoyama_2010}. The gap between the substrates is filled with a nematic liquid crystal and a vertical disclination arch spontaneously forms, pinned at its ends by the two surface defects, as shown schematically in Fig.~\ref{fig:experiment} (a). Viewed from above, the arch appears as a straight segment, as seen in a micrograph Fig.~\ref{fig:experiment} (b). At $0^{\circ}$ of twist, the length of the disclination line is approximately equal to the spacing between the two point defects which is $100\, \mu m$. The gap between the two substrates is approximately $30\, \mu m$.

An in-plane twist is then imposed between upper and lower substrates. After we increase the relative orientation angle, twisting the nematic director, the disclination bows and expands laterally into a horizontal arc as shown in Fig.~\ref{fig:experiment} (c-g). If the twist angle exceeds a threshold value,  the disclination arc expands enough to self-intersect and snap off a new loop, as shown in Fig.~\ref{fig:experiment} (h-i). The newly emitted disclination loop expands, leaving the original pinned segment intact. Experimental details are provided in Methods and in Ref.~\cite{y7_angelo_2017}.  A real time video of the experiment (Movie S1) is provided in Supplemental Information. 

The angle of $144^{\circ}$ was the smallest rotation for which we found loop emission within a reasonable waiting time, here about nine minutes.  If we stop at a slightly larger rotation angle, loop emission occurs within a shorter waiting time. For a rotation angle of $160^{\circ}$, the waiting time drops to zero.

This sequence of defect bowing, self-intersection, and loop snap-off is nearly identical to that observed for dislocation loop emission from a Frank-Read source in a crystalline solid; but the properties of dislocations in crystals and disclinations in nematic liquid crystals are not identical. Therefore, we perform simulations and theoretical analysis to explore the properties and dynamics of a Frank-Read source in a nematic.

\begin{figure}
\includegraphics[width=\columnwidth]{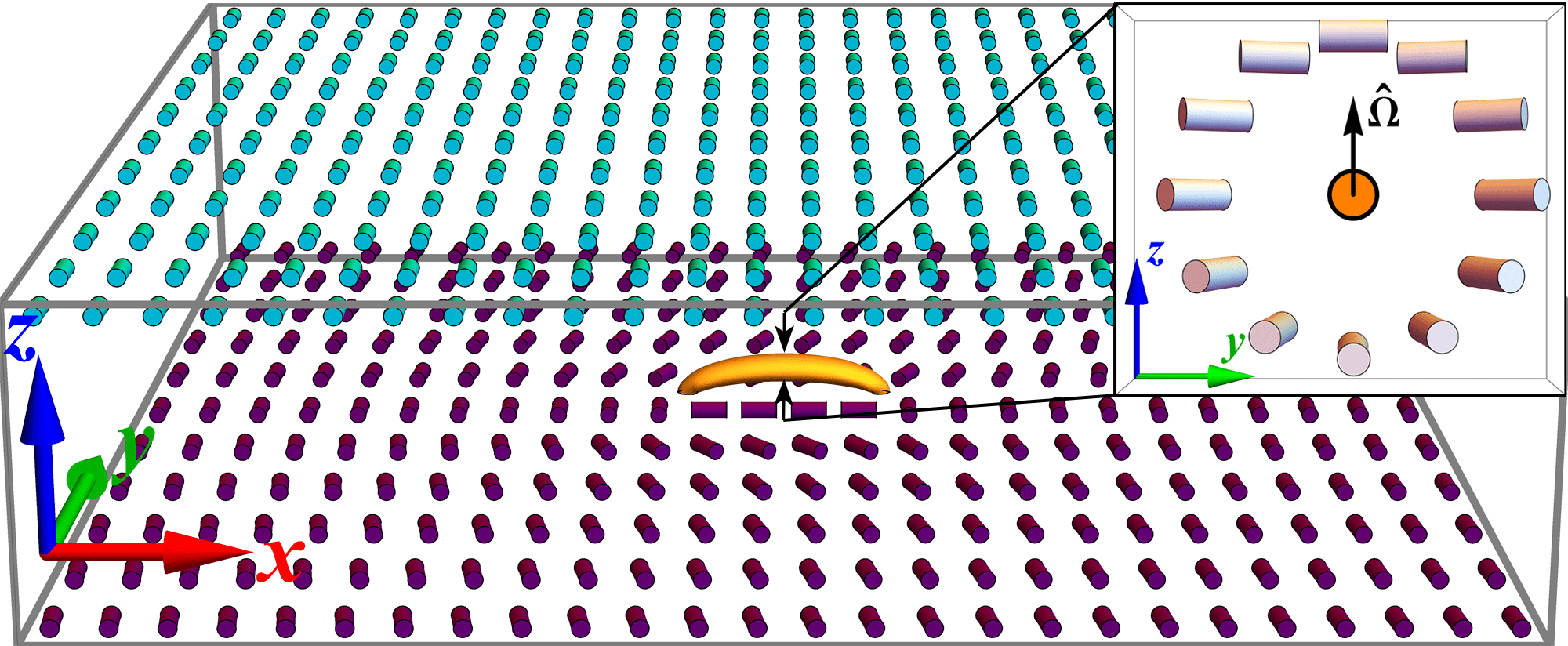}
\caption{Geometry of the simulation cell, for the order-tensor and the lattice simulations.  The bottom substrate (purple) has an anchoring pattern with two surface disclinations, while the top substrate (cyan) has uniform planar anchoring, which is initially aligned with the far-field of the bottom substrate.  A disclination arch (yellow) connects the two surface disclinations. Based on the director field around the disclination arch, the type of the disclination changes continuously from a $+1/2$ wedge disclination, to a twist disclination, eventually to a $-1/2$ wedge disclination along the line from left to right. The inset graph shows the cross section of the twist disclination in the disclination arch, and the $\hat{\Omega}$ vector indicates the rotation axis of the director around the disclination. In the example shown here, the arch is derived from order-tensor simulations.}
\label{fig:cellgeometry}
\end{figure}

\section{Simulations: Quasistatic Twist Deformation}

To gain insight into the Frank-Read mechanism in nematics, we perform numerical simulations of a nematic cell using the geometry shown in Fig.~\ref{fig:cellgeometry}, under conditions of quasi-static twist.  The bottom substrate has strong anchoring with two point defects of topological charges $+1/2$ and $-1/2$, similar (although not identical) to the experiment shown in Fig.~\ref{fig:experiment} \cite{Long2021,Guo2021,Afghah2018}.  The top surface has strong anchoring in a uniform planar direction.  The four side surfaces are free boundaries.  The spacing between surface disclinations is $w=4$ (arbitrary units), with cell size $120 \times 120 \times 10$. Inside the cell, orientational order is characterized by the nematic order tensor $\bm{Q}(\bm{r},t)$, with free energy given by a Landau-de Gennes series expansion.  We implement relaxational dynamics using an overdamped equation of motion, solved numerically via time-dependent finite-element modeling, using the software package COMSOL. Details are given in the Methods section.

\begin{figure*}[!ht]
\includegraphics[width=\textwidth]{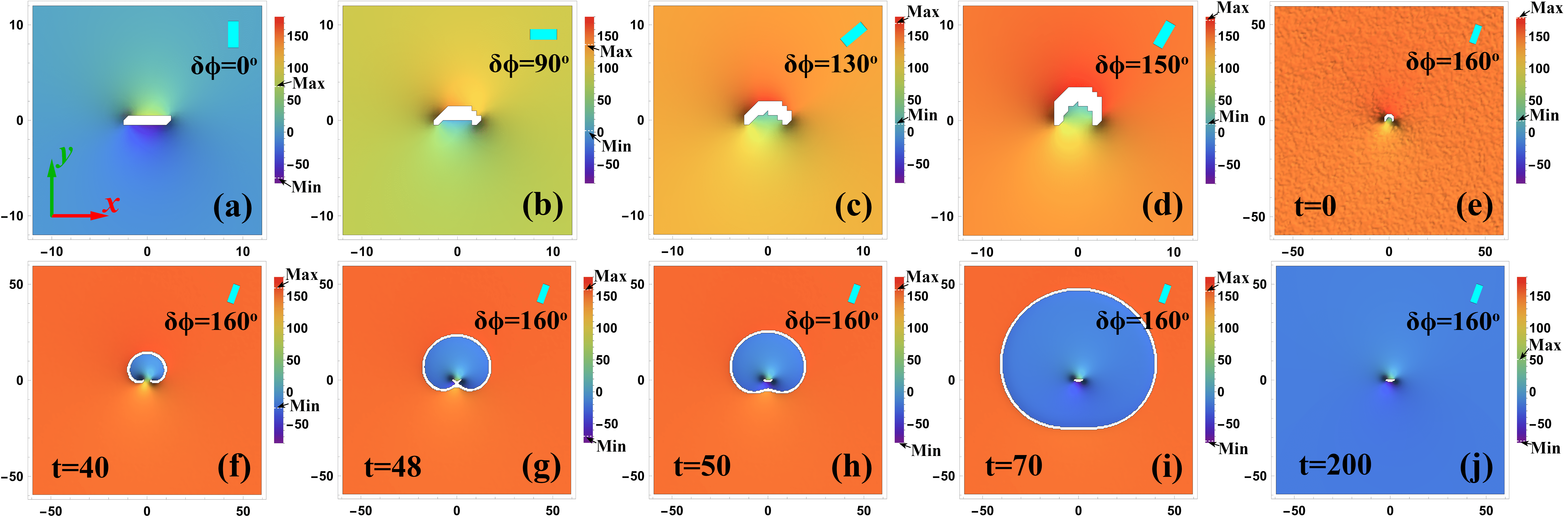}
\caption{Sequence of disclination structures as the top surface is rotated quasistatically, based on simulations of the nematic order tensor.  In these views from above, the color indicates the twist angle (in degrees) of the nematic director from the bottom to the top surface.  Inside the disclination cores, the twist angle is not uniquely defined because of biaxiality, and hence the color is shown as white.  (a)--(d) Steady states with $\delta\phi$ fixed at $0^{\circ}$, $90^{\circ}$, $130^{\circ}$, and $150^{\circ}$ respectively.  The size of the viewing window is $24\times 24$.  (e)--(j) Snapshots of the dynamic process when $\delta\phi$ is increased to $160^{\circ}$, which is slightly greater than the threshold angle.  The size of the viewing window is $120\times 120$. The cyan cylinder located at the top right corner of each image represents the orientation of the top surface.}
\label{fig:tensorsimulationtopviews}
\end{figure*}

Inside the cell shown in Fig.~\ref{fig:tensorsimulationtopviews}a, the top surface planar anchoring is initially aligned with the $y$ axis, which is also the far-field orientation of the bottom surface. We initialize the order tensor as a uniform nematic state aligned with the $y$ axis, and it relaxes into a stable configuration with a single disclination line connecting the two surface defects in an arch. Because the surface anchoring imposes parallel alignment with the two substrates, the resulting director field is also parallel to both substrates to minimize the elastic energy. As a result, the rotation axis of the director around the disclination ($\hat{\bm{\Omega}}$) is always perpendicular to the substrates, causing the disclination line to transform continuously from a $+1/2$ wedge disclination, to a twist disclination, and eventually to a $-1/2$ wedge disclination along its length.

Next we rotate the anchoring direction of the top surface through an angle $\delta\phi$. Loading is quasistatic; at each value of $\delta\phi$, we begin with the previous configuration of nematic order, and run the simulation until the nematic order relaxes to a new steady state.  As $\delta\phi$ increases from $0^{\circ}$ to $150^{\circ}$, the disclination bows laterally into a curve, as seen in Fig.~\ref{fig:tensorsimulationtopviews}a--d.  When $\delta\phi$ reaches $150^{\circ}$, the curve is approximately a semicircle.  The disclination separates regions with different twist angles between bottom and top substrates.  As discussed below, it is stabilized by the balance between the line tension (which favors a short disclination) and the twist energy (which favors moving the disclination to enlarge the low-twist region and reduce the high-twist region). We note that since the rotation of the top substrate does not introduce any $z$ component into the director field, the $\hat{\bm{\Omega}}$ vector of the bowed disclination line~\cite{Long2021} remains perpendicular to the substrates, implying that the topological feature of the bowed disclination line is still the same as the one for $\delta \phi = 0$.

As we increase applied twist, the system eventually reaches a threshold where the behavior changes radically. At $160^{\circ}$, the disclination moves through the series of shapes shown in Fig.~\ref{fig:tensorsimulationtopviews}e--j via overdamped relaxation.  Here the disclination does not find a stable or metastable position.  Instead, it expands; arms adjacent to the two fixed ends rotate toward each other, intersect, and merge (Fig.~\ref{fig:tensorsimulationtopviews}g), snapping off to form two new disclination lines (Fig.~\ref{fig:tensorsimulationtopviews}h).  One of these is a newly emitted pure twist disclination loop which expands (Fig.~\ref{fig:tensorsimulationtopviews}i) and eventually moves out of the simulation box by annihilating at the free boundaries (Fig.~\ref{fig:tensorsimulationtopviews}j).  Through emission, expansion, and escape of this loop, the net twist deformation in the cell is reduced by $\pi$ radians, which demonstrates a mechanism of stress release in nematic liquid crystals by generation and motion of topological defects.  The remaining disclination segment connects the two surface defects, and this source is ready to emit more loops under further imposed twist. A new loop is emitted for each additional $180^{\circ}$ of applied twist. An animation of the simulation (Movie S2) is provided in Supplementary Information. 

Because our model has mirror symmetry about the vertical plane crossing the two defect points on the bottom substrate at $\delta \phi=0$, and the nematic liquid crystal is achiral, we expect the same behavior for the pinned disclination line when $\delta \phi$ is negative, except that the pinned disclination line bows towards the opposite direction.

\section{Simulations: Finite Twist Rate and Thermal Effects}
To examine effects of finite twist rate, we simulate a nematic liquid crystal in a cell with the same geometry as Fig.~\ref{fig:cellgeometry}, using a Lebwohl-Lasher rotor model~\cite{Priezjev2001,Preeti2013,Skacej2021}. The liquid crystal is represented as a uniaxial nematic with a unit director field, $\hat{\bm{n}}_i$, defined on each site, $i$, of a cubic lattice, interacting through the nearest-neighbor potential $U_{ij}=-\epsilon P_2(\hat{\bm{n}}_i\cdot\hat{\bm{n}}_j)=-\epsilon[\frac{3}{2}(\hat{\bm{n}}_i\cdot\hat{\bm{n}}_j)^2-\frac{1}{2}]$. The director field evolves via overdamped rotor dynamics, relaxing along the potential energy gradient. Temperature is controlled via a Langevin thermostat. We implement the model for fast execution using graphics processing unit (GPU) acceleration. Model details are provided in Methods.

\begin{figure*}[ht]
\includegraphics[width=.9\textwidth]{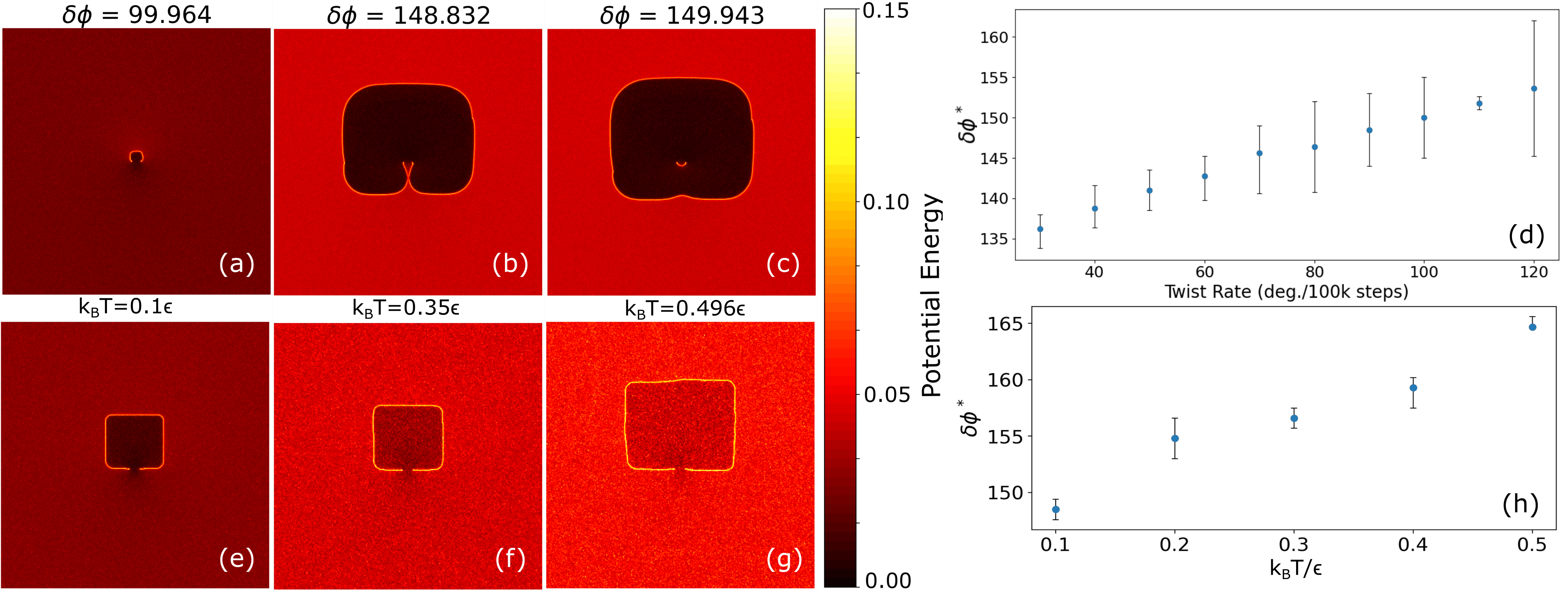} 
\caption{Simulations of Frank-Read source behavior at finite temperature, via the Lebwohl-Lasher model, with defect spacing $w=16$ and cell thickness $d=10$. (a-c) Disclination loop emission under continuous twist of the planar anchoring orientation at the top surface, with rate $90^{\circ}/100k$ time steps, and temperature $k_B T=0.1\epsilon$. (d) Threshold twist angle
$\delta \phi^*$
for loop emission increases as a function of applied twist rate.
(e-g) Disclination loop with fixed sub-critical applied twist undergoes stable expansion as temperature increases, due to changes in Frank elasticity and line tension. (h) Threshold twist angle $\delta \phi^*$  for loop emission increases as a function of temperature.}
\label{fig:latticesimulation}
\end{figure*}

Figure~\ref{fig:latticesimulation}(a-c) shows disclination loop emission at finite temperature and under continuously increase twist angle; color indicates local potential energy density. As twist increases, the disclination bows and expands into an arc.  At a twist angle $\delta \phi^*$, the arc self-intersects (Fig.~\ref{fig:latticesimulation}b), snapping off to form a new loop which expands and moves out of the simulated cell via annihilation at the free surfaces (Fig.~\ref{fig:latticesimulation}c). We note that the Lebwohl-Lasher model introduces anisotropic line tension which causes curved disclinations to facet, an artifact of the underlying cubic lattice. An animation is shown in Supplementary Information (Movie S3).

We vary the applied twist rate of the top surface anchoring orientation. As shown in Fig.~\ref{fig:latticesimulation}d, the threshold for loop emission  $\delta \phi^*$ (identified here as the angle corresponding to peak potential energy) rises with increasing twist rate. This finding is in qualitative agreement with the behavior of Frank-Read sources in crystalline solids, where the strain threshold for loop emission typically rises with increasing strain rate~\cite{Gurrutxaga2015}. 

Next we model thermal effects. We performed simulations at different  temperatures and find that the threshold angle for loop emission rises with temperature (Fig.~\ref{fig:latticesimulation}h). This increase in strain at loop emission as a function of temperature is qualitatively similar to behavior of Frank-Read sources in crystalline solids ~\cite{Geipel1996}. 

As another measure of thermal effects, we study the behavior of a bowed disclination segment under gradually increasing temperature. We hold the twist angle at a fixed, subcritical value where the bowed disclination segment is metastable. As temperature rises from $k_B T=0.1\epsilon$ to $0.5\epsilon$, the disclination arc expands outwards (Fig.~\ref{fig:latticesimulation}e-g).  This spontaneous expansion is similar to the disclination ``blowout" at the nematic-isotropic transition, reported in previous Lebwohl-Lasher simulations~\cite{Priezjev2001}, but occurs within the nematic phase. It is also consistent with the change in disclination shape on heating observed experimentally by Modin et al~\cite{Modin2023}.

We note that the Lebwohl-Lasher model only takes account of director relaxation without hydrodynamics. In future work on finite strain rate studies, we will add hydrodynamics ~\cite{Yeomans2001,Copar2021,Duclos2020,Mandal2019} to investigate backflow effects in loop emission and resulting rheological response.

\section{Analytic Theory: 2D Free Energy}

\begin{figure}
\includegraphics[width=\columnwidth]{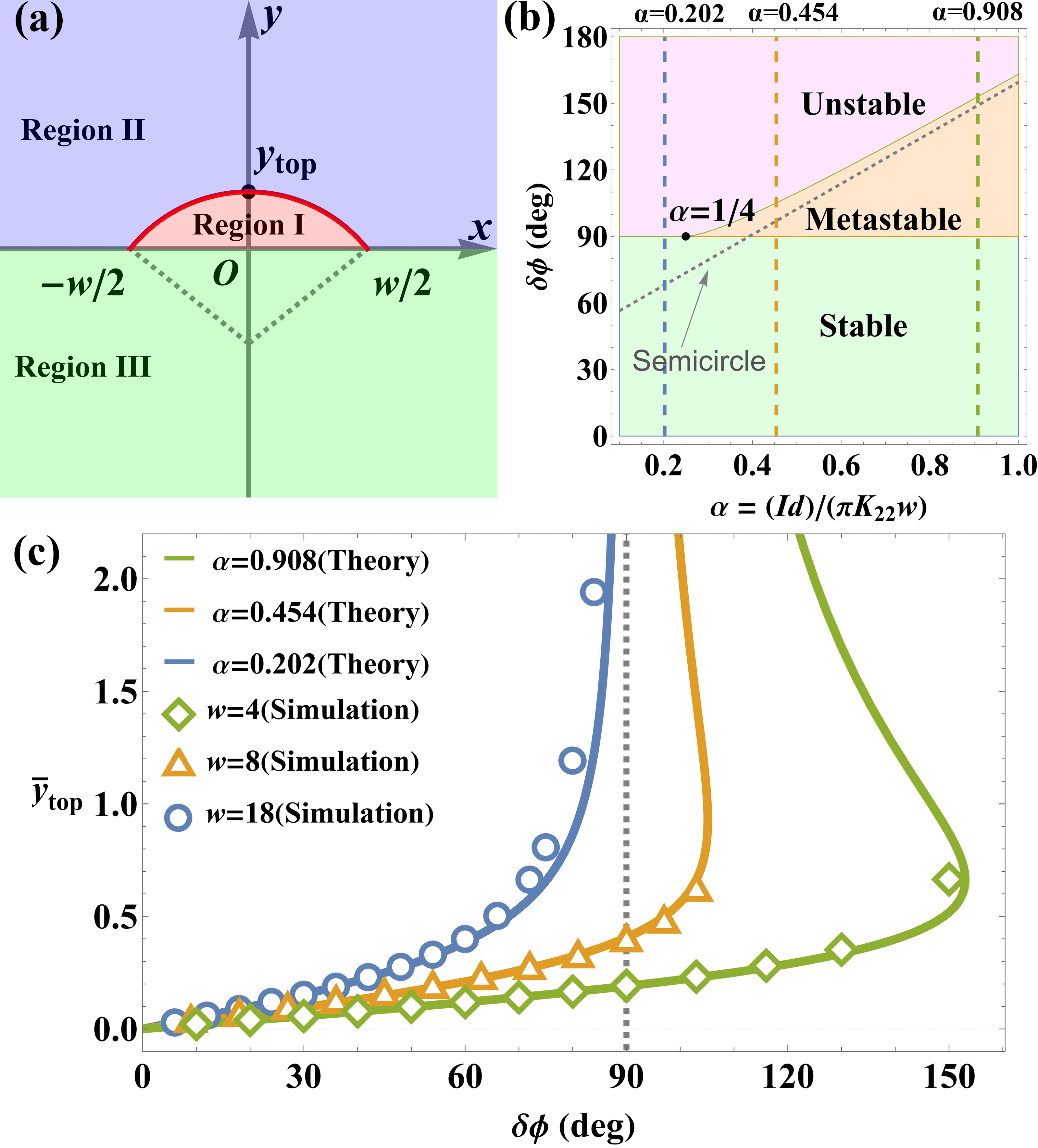}
\caption{(a) Geometry for the theoretical free energy of the Frank-Read source.  The thick red line represents the curved disclination, viewed from above.  (b) Stability diagram in terms of angle $\delta\phi$ and dimensionless parameter $\alpha$, indicating whether the curved disclination is stable, metastable, or unstable (so that it must grow to infinity).  Along the gray dashed line, the shape is a semicircle.  (c) Comparison of theoretical predictions with nematic-order-tensor simulation results, for different defect spacings $w$.}
\label{fig:theory}
\end{figure}

For an analytic theory of the Frank-Read source, we consider the geometry in Fig.~\ref{fig:theory}a.  We assume the nematic director lies in the $(x,y)$ plane so that $\hat{\bm{n}}=(\cos{\phi},\sin{\phi},0)$.  On the lower substrate, it is anchored at orientation $\phi_B(x,y)$, given in Methods, which has surface defects at $(\pm w/2,0)$.  On the top substrate, it is anchored at $\phi_T=(\pi/2)+\delta\phi$, with $\delta\phi\ge0$, independent of $x$ and $y$.  This geometry induces formation of a disclination shown by the thick red line in Fig.~\ref{fig:theory}a.

We approximate the disclination shape as a circular arc between the surface defects, passing through $(0,y_\text{top})$.  For $y_\text{top}=0$, this shape is a straight line between the two surface defects.  For $y_\text{top}=w/2$, it is a semicircle connecting the surface defects.  For $y_\text{top}\to+\infty$, it is an almost-complete circle in the positive $y$ region, just touching the surface defects.

The free energy of this configuration has two components.  One component is the line energy, proportional to the length of the disclination.  The second component is the elastic energy associated with director twist between the two substrates.  As $y_\text{top}$ increases and the disclination grows, the line energy increases but the elastic energy decreases, because more area is transformed from high twist to reduced twist.  The balance between these two competing terms is worked out in detail in Methods.  The minimum free energy occurs at
\begin{equation}
\delta\phi=\arctan(2\bar{y}_\text{top})+\frac{8\alpha \bar{y}_\text{top}}{1+4\bar{y}_\text{top}^2},
\label{eq:DeltaPhiAndycBar}
\end{equation}
which implicitly shows how disclination shape evolves with $\delta\phi$. Here, $\bar{y}_\text{top} = y_\text{top}/w$ is a dimensionless parameter characterizing $y_\text{top}$ normalized by defect separation $w$.  Likewise, $\alpha=(I d)/(\pi K_{22}w)$ is a dimensionless parameter representing line tension $I$ times cell thickness $d$, compared with twist elastic constant $K_{22}$ and defect separation $w$.

By analyzing Eq.~(\ref{eq:DeltaPhiAndycBar}), we identify distinct regimes of stability, shown in Fig.~\ref{fig:theory}b.  First, consider the case of $\alpha<1/4$.  As $\delta\phi$ rotates from $0$ to $\pi/2$, $y_\text{top}$ increases from $0$ to $+\infty$.  Hence, the disclination grows from a straight segment at $\delta\phi=0$ into a semicircle, then into an almost-complete circle connecting the surface defects at $\delta\phi=\pi/2$.  All of these shapes are stable minima of the free energy.  If $\delta\phi$ rotates beyond $\pi/2$, then there is no stable shape in the positive $y$ region.  Rather, the disclination must grow to infinity.

If $\alpha>1/4$, the behavior is different.  As $\delta\phi$ rotates from $0$ to $\pi/2$, $y_\text{top}$ increases from $0$ to a finite value, which depends on $\alpha$.  If $\delta\phi$ rotates beyond $\pi/2$, then $y_\text{top}$ continues to increase, and the disclination grows larger in the positive $y$ region as a \emph{metastable} structure (while the \emph{stable} structure is a disclination in the negative $y$ region).  The metastable disclination can grow up to
\begin{align}
\bar{y}_\text{top}^\text{max}&=\frac{1}{2}\sqrt{\frac{4\alpha+1}{4\alpha-1}},\\
\delta\phi^\text{max}&=\frac{\sqrt{16\alpha^2-1}}{2}+\arctan\sqrt{\frac{4\alpha+1}{4\alpha-1}}.\nonumber
\end{align}
If $\delta\phi$ rotates beyond $\delta\phi^\text{max}$, the disclination becomes \emph{unstable;} it finds no steady state in the positive $y$ region.  Instead, it must expand to infinity.

Predictions of this analytic theory can be compared with the quasistatic nematic-order-tensor simulation presented above.  Figure~\ref{fig:theory}c shows simulation results for defect separation $w=4$, $8$, and $18$.  For each value of $\delta\phi$, we analyze the simulated shape to extract $y_\text{top}$.  These results can be fit to Eq.~(\ref{eq:DeltaPhiAndycBar}), with the dimensionless parameter $\alpha=0.908$, $0.454$, and $0.202$.  The prediction and simulations agree well over the full range of $\delta\phi$, and they give consistent values for the limit of metastability $\delta\phi^\text{max}$.  As expected, $\alpha$ is inversely proportional to $w$.  From the values of $\alpha$ and $w$, along with the cell thickness $d=10$ and twist elastic constant $K_{22}=3.0$, we extract the line tension $I=3.4$ (all in the arbitrary units defined in Methods). Line tension is assumed to remain constant throughout the dynamic expansion process. This value is reasonable in terms of Landau-de Gennes theory.

The predictions can be used to interpret the experiment presented in Fig.~\ref{fig:experiment} and Supplementary Video~S1. Because the disclination line expands rapidly past $\delta\phi=144^\circ$, we estimate the dimensionless parameter $\alpha=(I d)/(\pi K_{22}w)=0.9$. Because the surface defect spacing is $w=100$~$\mu$m, and the cell gap is $d=30$~$\mu$m, we have the ratio $I/K_{22}=10$. For example, the twist elastic constant might be $K_{22}=4$~pN, and disclination line tension $I=40$~pN. These numerical estimates can be compared with a recent experimental and theoretical study by Modin et al ~\cite{Modin2023}. Our dimensionless parameter $\alpha$ is equivalent to their parameter $2\tilde{\gamma}$, and our estimate $I/K_{22}=10$ is similar to their reported ratio of 12 to 18, depending on temperature.

\section{Analytic Theory: Peach-Koehler Force}

In crystalline solids, the standard theory of a Frank-Read source is based on the balance between the Peach-Koehler force (which pushes the dislocation forward) and the dislocation line tension (which pulls it back). The Peach-Koehler force is a force per length acting on dislocations in crystalline solids under stress~\cite{Peach1950,Weertman1964,Anderson2017}. It can be written as $\bm{f}_\text{PK}=(\bm{b}\cdot\bm{\sigma})\times\hat{\bm{t}}$, where $\bm{b}$ is the Burgers vector, $\bm{\sigma}$ is the local stress tensor, and $\hat{\bm{t}}$ is the local tangent vector of the dislocation. An estimate of its magnitude is $f_\text{PK}=b\sigma$. By contrast, the line tension provides a force per length of $I/r$, where $I$ is the dislocation energy per length and $r$ is the local curvature radius.  These forces balance when $b\sigma=I/r$. As an estimate, the dislocation energy per length is $I\approx G b^2$, where $G$ is the shear modulus of a solid. Furthermore, the minimum curvature radius is $r\approx w/2$, where $w$ is the spacing between the pinned defects. Hence, the curved dislocation is only stable up to a maximum stress of $\sigma\approx 2 G b/w$. Beyond that stress, the Frank-Read source becomes active and produces dislocation loops.

We can apply the same argument to the Frank-Read source in a nematic liquid crystal. Early work by Kl\'eman~\cite{Kleman1983} showed that nematic disclinations experience a Peach-Koehler force, but this effect has not been studied much in the liquid-crystal literature. In a recent paper~\cite{Long2021}, our group developed a theory for this force as
\begin{equation}
\bm{f}_\text{PK}=\left(\pi\hat{\bm{\Omega}}\cdot\bm{\sigma}^\text{eff}\right)\times\hat{\bm{t}}.
\label{fPK}
\end{equation}
Here, $\hat{\bm{\Omega}}$ is the rotation vector of the director field around the disclination, with $\pi\hat{\bm{\Omega}}$ analogous to a Burgers vector. The effective stress tensor $\bm{\sigma}^\text{eff}$ is the difference of torques between the top and bottom surfaces, normalized by the area over which the torques are applied. Hence, a simple estimate of the Peach-Koehler force per length is $f_\text{PK}=\pi\sigma^\text{eff}$. It competes with the disclination line tension, which provides a force per length of $I/r$, and the forces balance when $\pi\sigma^\text{eff}=I/r$. The disclination energy per length is approximately $I\approx K\pi^2$, assuming all Frank elastic constants are equal to $K$. The minimum curvature radius is $r\approx w/2$, is half the spacing between the surface defects.

Hence, the curved disclination is only stable up to the maximum effective stress of $\sigma^\text{eff}\approx 2 K\pi/w$. Beyond that stress, it produces disclination loops. Critical stress for dislocation loop emission in crystalline solids also scales inversely with the distance between pinning points.

To make our theory more precise, we return to Eq.~(\ref{fPK}). The tangent vector $\hat{\bm{t}}$ runs along the disclination.  It is in the $(x,y)$ plane, except in small regions near the two surface defects.  The rotation vector $\hat{\bm{\Omega}}$ is vertical everywhere.  We must make an arbitrary choice for the signs of $\hat{\bm{t}}$ and $\hat{\bm{\Omega}}$, and these two signs must be compatible with each other.  Let us choose that $\hat{\bm{t}}$ goes from the $+1/2$ to the $-1/2$ surface defect, and  $\hat{\bm{\Omega}}=+\hat{\bm{z}}$.  From Ref.~\cite{Long2021}, assuming equal Frank elastic constants, the effective stress tensor is $\sigma^\text{eff}_{ji}=K(\hat{\bm{n}}\times\partial_i\hat{\bm{n}})_j$.  As a simple approximation, we consider only the variation of $\hat{\bm{n}}$ in the $z$ direction.  The effective stress of rotating the top alignment from $\phi_T=\pi/2$ to $\phi_T=(\pi/2)+\delta\phi$ then becomes $\sigma^\text{eff}_{zz}=K(\delta\phi-\phi_B)/d$, with all other components zero.  Hence, the Peach-Koehler force per length acting on the disclination is
\begin{equation}
\bm{f}_\text{PK}=\pi K\frac{\delta\phi-\phi_B}{d}\hat{\bm{z}}\times\hat{\bm{t}}.
\end{equation}
In our geometry, $\hat{\bm{t}}$ is mostly in the $+\hat{\bm{x}}$ direction, and hence $\bm{f}_\text{PK}$ is mostly in the $+\hat{\bm{y}}$ direction.  It pushes the disclination toward the positive $y$ region, as seen in both experiment and simulations.  By comparison, line tension provides a force per length of 
\begin{equation}
\bm{f}_\text{tension}=-\frac{I}{r}\hat{\bm{z}}\times\hat{\bm{t}},
\end{equation}
with direction inward toward the center of curvature.  By balancing these forces, we predict the curvature radius
\begin{equation}
r=\frac{I d}{\pi K(\delta\phi-\phi_B)}.
\end{equation}
Using expressions for $r$ and alignment angle $\phi_B$ from Methods, we see that this equation is exactly equivalent to Eq.~(\ref{eq:DeltaPhiAndycBar}) derived from free energy minimization.  It predicts the maximum stable stress of $\sigma^\text{eff}_{zz}=(2I)/(\pi w)$, consistent with the estimate above, in the limit of $\alpha=(I d)/(\pi K w)\gg1$.

One further benefit of the Peach-Koehler force concept is that it provides a way to estimate the velocity of an expanding disclination loop.  In general, a disclination moves with overdamped dynamics, following the equation of motion $\bm{f}_\text{total}=\zeta\bm{v}$, where $\bm{f}_\text{total}=\bm{f}_\text{PK}+\bm{f}_\text{tension}$ is the total force (per length), $\zeta$ is the drag coefficient (per length), and $\bm{v}$ is the velocity.  In Methods, we use this dynamic equation to determine the velocity of a large expanding disclination loop.  We find $v=0.81(\delta\phi-\pi/2)$, in the arbitrary units.  For example, $\delta\phi=160^\circ$ gives $v=1.00$, and $\delta\phi=108^\circ$ gives $v=0.26$.

\begin{figure}
\includegraphics[width=\columnwidth]{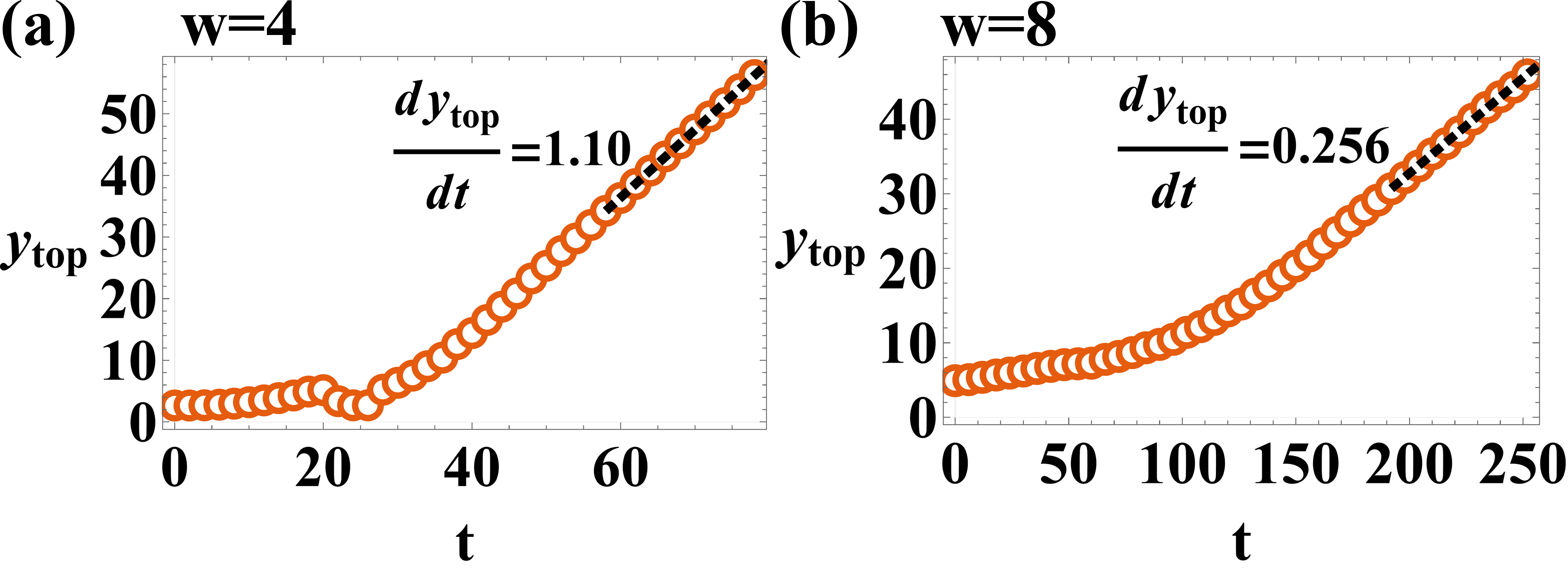}
\caption{Motion of $y_\text{top}$ during the unstable expansion process, from nematic-order-tensor simulations.  (a)~Defect spacing $w=4$ and angle $\delta\phi=160^\circ$.  (b)~Defect spacing $w=8$ and angle $\delta\phi=108^\circ$.  The black dashed lines show linear fits in the limit of $y_\text{top}\gg w$.}
\label{fig:dynamics}
\end{figure}

To check this estimate, we monitor loop expansion after $\delta\phi$ exceeds the critical threshold, as in Fig.~\ref{fig:tensorsimulationtopviews}e--j, tracking $y_\text{top}$ as a function of time.  Figure~\ref{fig:dynamics}a shows the motion in the simulation with defect spacing $w=4$ at angle $\delta\phi=160^\circ$.  After an initial phase of slow growth, presumably limited by line tension, $y_\text{top}$ increases at the steady velocity of 1.10 arbitrary units.  Figure~\ref{fig:dynamics}b shows corresponding results in the simulation with $w=8$ at $\delta\phi=108^\circ$.  In this case, $y_\text{top}$ increases at the steady velocity of 0.256 arbitrary units.  These simulation results are quite consistent with the prediction.

We note that the present analysis using the Peach-Kohler force was carried out under the approximation of equal splay and bend Frank elastic constants. In this limit we find that the analogy to the Peach-Kohler mechanism in crystalline solids works well. The case of unequal Frank constants is more challenging and will be the subject of future investigation.

\section{Discussion and Outlook}

In closely related experimental and theoretical work, Jiang et al~\cite{Jiang2022} studied a lattice of surface-pinned disclination segments in a nematic cell. Under director twist imposed by optically rewriting surface anchoring orientation on the opposite substrate, the defects underwent programmable transformations via expansion, merging, and reconfiguration. We interpret this geometry as an array of Frank-Read sources that are closely spaced, such that expanding disclination segments merge and reconnect, pre-empting loop emission. We predict that spacing the pinned segments further apart will enable emission of expanding loops, forming an array of Frank-Read sources. Expanding loops will then collide and reconnect. Independent of defect spacing, this mechanism creates a localized slip plane for twist deformation, in analogy to slip planes in metals. We speculate that this technique could be implemented to control microstructural evolution and rheological response (e.g. lubrication) for flow of nematics under shear between patterned substrates.

The Frank-Read mechanism could arise more generally in nematics via other stimuli that deform the director field, such as fluid flow, electric or magnetic fields, or by internal stresses in active nematics. In nematic flow through a microfluidic channel~\cite{Salamon2022,Sengupta2011,Liu2019, Copar2021} or in a Couette cell ~\cite{Mather1996}, disclination nucleation has been observed at confining walls due to the presence of surface irregularities which pin disclinations.

In nematics, we have so far only considered Frank-Read sources with pinning points located on a confining surface. In crystals, by contrast, dislocations can pin at random along their length via jog formation, interaction with impurities, or entanglement~\cite{Anderson2017, Gurrutxaga2015}, generating new Frank-Read sources.  Disclinations in pure nematics do not spontaneously pin by any analogous mechanism. However addition of colloid particles can induce disclinations to form knots and complex entangled networks~\cite{Copar2021}, suggesting that a nematic-colloid composite might demonstrate spontaneous Frank-Read source formation under imposed flow. We speculate that resulting microstructural evolution might show complex behaviors in analogy to mechanisms governing plasticity in crystalline solids. On the other hand, thanks to the interaction between colloids and disclinations, the Frank-Read source also provides a controllable approach for redistribution of colloids immersed in nematic liquid crystals responding to external stimuli. From recent studies on manipulation of light using nematic disclinations~\cite{loussert2013manipulating,vcanvcula2014generation,meng2023topological}, we speculate that disclinations controlled and created via the Frank-Read mechanism might find use in novel optical applications.

In 3D active nematics, active stress drives both fluid flow and director rotation. Studies by Duclos et al.~\cite{Duclos2020} revealed 3D details of microstructural evolution. They showed that colliding disclinations merge and reconnect without forming long-lived pinning points. While they observed that disclination self-intersection can generate new loops, in the absence of defect pinning, no Frank-Read source formation was observed. In future work, we will investigate whether a network of Frank-Read sources can be created in an active nematic by inscribing defect pinning points on confining substrates. 

It is also interesting to note that the homogeneous defect loop nucleation mechanism in active nematics, as shown in Duclos et al.~\cite{Duclos2020}, is not typically observed in plastic deformation of ductile crystals. Dislocation loop generation from Frank-Read sources occurs at far lower critical stress, and thus heterogeneous defect nucleation preempts homogeneous nucleation. Reasoning by analogy, we speculate that a high density network of Frank-Read sources in an active nematic could produce a more ordered pattern of defect generation and flow, with stresses never reaching the threshold for homogeneous loop nucleation. Any mechanism that produces disclination pinning--e.g. surface defects, colloids, or micropillars --could thus modify pattern formation and rheology of both active and passive nematics.

In conclusion, we have presented experimental, theoretical, and simulation studies of the Frank-Read mechanism in nematic liquid crystals. The experimental micrograph in Figure~\ref{fig:experiment}(b-i) shows the signature shape evolution of a Frank-Read source.  Quasistatic simulations using a nematic order tensor and Landau-de Gennes free energy demonstrate the mechanism. Finite strain rate simulations via the Lebwohl-Lasher rotor model show that threshold twist for loop emission rises with increasing strain rate, analogous to behavior in metals. We also observe spontaneous loop expansion under increase of temperature. We develop analytical theory of loop emission by minimizing a 2D free energy, and also via balance of Peach-Koehler and line tension forces, showing close agreement with quasistatic simulations. 

As a historical aside, we note that Sir Frederick Charles Frank made crucial contributions to the study of both crystal plasticity and nematic liquid crystals. He co-discovered the Frank-Read source mechanism together with Thornton Read in 1950 ~\cite{Frank1950}; and also formulated the Frank Free energy for liquid crystals with contributions from splay, twist, and bend, in 1958 ~\cite{Frank1958}. The work described here brings together and builds upon Frank's contributions to both fields.

\section{Methods}
\subsection{Dynamic Cell Experiment}
The dynamic cell \cite{y6_culbreath_2015,y7_angelo_2017} is comprised of two glass substrates, top and bottom, with a nematic liquid crystal (in this case 5CB) sandwiched in between (Fig.~\ref{fig:experiment}a). The bottom substrate is fixed on a rotating stage whose rotation axis is along the substrate normal (z-axis). The top substrate is fixed on a stage which translates along the z-axis. This setup allows the bottom substrate to twist with respect to the top substrate, and for the top substrate to translate to change the cell gap between substrates. The cell gap is measured using a capacitance sensor.

The bottom substrate is a 1 inch $\times$ 2 inch glass substrate.  We used a plano-convex lens (1 inch diameter; with a radius of curvature of two meters) for the top substrate. The top substrate is coated with PI-2555 polyimide and is rubbed 10 times with a velvet cloth and marked to indicate the rubbing direction. The bottom substrate is coated with an azo-based photoalignment compound (SD-1) that can be photoaligned by polarized UV light \cite{y9_chigrinov_2008}.

Photoalignment of the bottom substrate is done in two steps. First, the entire substrate is exposed to polarized UV light ($8 \,mW/cm^2$ for 1 minute) to induce uniform planar anchoring. Surface defects are then generated in the surface anchoring pattern using a maskless photoalignment system\cite{y6_culbreath_2015,y8_culbreath_2011,y10_glazar_2015}.

 The patterned anchoring is comprised of four regions with different planar alignment angles (Fig.~\ref{fig:experiment}a).  The central rectangular region $(100 \,\mu m \times 50 \,\mu m)$ is split into three segments as shown in Fig.~\ref{fig:experiment}a. The maskless system uses a rotating polarizer to expose the substrate with polarized UV light. Each region is aligned in turn: first the polarizer is rotated to generate the desired alignment angle. Then the maskless system creates an image of the region focused on the substrate surface.

Once the substrate preparation is complete, the dynamic cell is assembled. First a small amount of 5CB is deposited on the bottom substrate surface. The disclination line (Fig.~\ref{fig:experiment}b) spontaneously forms between the two surface defects. Then the top substrate is translated downwards, put in contact with the liquid crystal, and is further translated downward until a cell gap of approximately $30 \,\mu m$ is attained.

\subsection{Simulations of nematic order tensor}

Nematic order is represented by the tensor $\bm{Q}(\bm{r},t)$, which is a function of position and time.  In uniaxial regions, outside of disclination cores, this tensor is related to the nematic director $\hat{\bm{n}}$ and scalar order parameter $S$ by
$Q_{ij}=S(\frac{3}{2}n_i n_j-\frac{1}{2}\delta_{ij})$.  The free energy is given by the standard Landau-de Gennes series expansion
\begin{align}
F &= \int d^3 r \biggl[\frac{A}{2}\Tr\bm{Q}^2+\frac{B}{3}\Tr\bm{Q}^3+\frac{C}{4}\left(\Tr\bm{Q}^2\right)^2\nonumber\\
&\qquad\qquad+\frac{L}{2}(\partial_i Q_{jk})(\partial_i Q_{jk})\biggr],
\label{free_energy_simulation}
\end{align}
where the first three terms are the bulk free energy, and the last term is the distortion free energy. We choose the coefficients $A=-1$, $B=-12.3$, $C=10$, and $L=2.32$ in arbitrary units.  The ratios of these coefficients correspond to the ratios in the liquid crystal 5CB, provided that the unit length in our simulations corresponds to 4.45 nm.  With these coefficients, the bulk scalar order parameter is the minimum of the bulk free energy $S=0.53$.  The single elastic constant $L$ for the nematic order tensor implies a single Frank elastic constant for the director $K=\frac{9}{2}S^2 L=3.0$.

To model pure relaxational dynamics, we use the overdamped equation of motion for $\bm{Q}(\bm{r},t)$,
\begin{equation}
\Gamma\frac{\partial Q_{ij}}{\partial t}=-\frac{\delta F}{\delta Q_{ij}},
\end{equation}
where $\Gamma$ is the rotational viscosity coefficient in the tensor representation.  We choose units of time such that $\Gamma=1$.

In the dynamic evolution, we impose strong planar anchoring conditions on the bottom and top surfaces, with $\hat{\bm{n}}=(\cos\phi,\sin\phi,0)$.  On the bottom, the director configuration is
\begin{equation}
\phi_B (x,y)=-\frac{1}{2}\arctan{\frac{x+w/2}{y}} + \frac{1}{2}\arctan{\frac{x-w/2}{y}} + \frac{\pi}{2}.
\label{eq:phiB}
\end{equation}
This pattern has a $+1/2$ defect at $(-w/2,0)$ and a $-1/2$ defect at $(+w/2,0)$, which enables a pinned disclination line connecting the two defects through the interior of the liquid-crystal cell.  On the top, the director has the uniform alignment
\begin{equation}
\phi_T =\frac{\pi}{2}+\delta\phi.
\label{eq:phiT}
\end{equation}
The alignment of the top surface is initially along the $y$ direction, and $\delta\phi$ is the angle by which it is rotated from the initial state.

Disclinations are visualized by the biaxiality parameter $\beta=1-6(\Tr\bm{Q}^3)^2/(\Tr\bm{Q}^2)^3$.  In particular, Fig.~\ref{fig:cellgeometry} shows the isosurface of $\beta=0.7$.  The thickness of the disclination corresponds to the core radius, which can be estimated from Landau-de Gennes theory as
\begin{equation}
r_\text{core}=\frac{1}{2}\sqrt{\frac{L}{|\frac{3}{4}A S^2+\frac{1}{4}B S^3+\frac{9}{16}C S^4|}}=1.6
\end{equation}
in the arbitrary units.  We recognize that this value is unrealistically large, in comparison with the defect spacing and system size.  By exaggerating this parameter, we reduce the ratio of length scales, which allows the simulations to run more quickly.  That is a drawback of simulations of the nematic order tensor, which is addressed in simulations of the lattice model.

\subsection{Simulations of lattice model}
The Lebwohl-Lasher model is a simplified version of the order-tensor representation of a nematic liquid crystal, where we assume equal elastic constants, no biaxiality, and scalar order parameter of uniform magnitude.  Spins on a $N\times N\times N$ cubic lattice represent the director field of a nematic liquid crystal. These spins can rotate freely, but they do not have any translational motion~\cite{Lebwohl1972,Pasini2000}. The Lebwohl-Lasher model is effective at simulating large systems on the order of $10^6$ spins, where each spin could represent a region of similarly aligned molecules~\cite{Pasini2000,Priezjev2001}. Spins interact through the potential
\begin{equation}
H=-\epsilon\sum_{\langle i,j\rangle}P_2(\hat{\bm{n}}_i\cdot\hat{\bm{n}}_j)
=-\epsilon\sum_{\langle i,j\rangle}\left[\frac{3}{2}(\hat{\bm{n}}_i\cdot\hat{\bm{n}}_j)^2 -\frac{1}{2}\right],
\end{equation}
where $i$ and $j$ are neighboring lattice sites, and $\epsilon$ is a positive interaction constant~\cite{Pasini2000,Skacej2021}.

The Lebwohl-Lasher potential effectively brings neighboring spins into parallel alignment and accurately reproduces disclination behavior of liquid crystal systems~\cite{Pasini2000,Priezjev2001,Preeti2013}.  Spins can be fixed at the boundary and patterned to represent a strongly anchored layer.  In this study, we use the anchoring patterns of Eqs.~(\ref{eq:phiB}--\ref{eq:phiT}) on the bottom and top surfaces, respectively, just as in the simulations of the nematic order tensor.

Many Lebwohl-Lasher models in the literature use a stochastic Monte-Carlo algorithm to study equilibrium behavior at finite temperature~\cite{Priezjev2001,Pasini2000,Skacej2021}.  However, we use an overdamped torque relationship between the neighboring spins to minimize our energy, creating a Lebwohl-Lasher rotor model~\cite{Pasini2000,Skacej2021}.  This dynamical system allows us to study microstructural evolution where the system continuously moves towards the lowest energy state.  The torque between neighboring spins is calculated from the derivative of the energy with respect to orientation,
\begin{equation}
\bm{\tau}_\text{bond}=\epsilon(\hat{\bm{n}}_i\times\hat{\bm{n}}_j)(\hat{\bm{n}}_i\cdot\hat{\bm{n}}_j).
\end{equation}
To allow the system to overcome energy barriers at constant temperature and reach the global energy minimum, we also add another term to the torque associated with an angular-momentum-conserving Langevin thermostat scaled with a temperature parameter $\eta= \sqrt{8 k_b T dt}$~\cite{Loft1987}.  This torque is then applied equally and oppositely to the neighboring spins $\hat{\bm{n}}_i$ and $\hat{\bm{n}}_j$.

We set each spin's angular velocity proportional to its total torque, resulting in overdamped rotation.  The angular velocity then becomes $\bm{\omega}=C\bm{\tau}$, where $C$ is a mobility factor inversely proportional to the rotational viscosity.  Hence, the spins in the Lebwohl-Lasher model are updated by
\begin{equation}
\frac{\hat{\bm{n}}_i(t+dt) - \hat{\bm{n}}_i(t)}{dt} =\bm{\omega}_i(t)\times\bm{n}_i(t)
\end{equation}
in each time step $dt$.

We note that the underlying cubic lattice in the 3D Lebwohl-Lasher model gives rise to anisotropy in disclination line tension. The lattice favors disclinations running in orthogonal directions, leading to rectangular shapes. Additional interactions with the edges of the square simulation box enhance the rectangular shape of the disclination when it is allowed to completely relax. This rectangular shape is not as pronounced when under a constant twist rate. Furthermore, the motion of a disclination requires overcoming a small energy barrier between lattice sites, in analogy to the Peierls-Nabarro barrier for dislocations in crystals~\cite{Selinger2011}.
\subsection{Minimization of 2D free energy}

We consider a nematic liquid crystal in the geometry shown in Fig.~\ref{fig:theory}a.  The bottom and top substrates have strong anchoring with the configurations given in Eqs.~(\ref{eq:phiB}--\ref{eq:phiT}), respectively.  Hence, the bottom substrate has two surface defects at positions $(\pm w/2,0)$, and these defects are connected by a disclination line.  We model this disclination line by a circular arc passing through $(0,y_\text{top})$.  The total free energy is the sum of two terms:  the line tension associated with the disclination, and the twist free energy of the director field between the bottom and top substrates.  We wish to minimize the total free energy over the parameter $y_\text{top}$.

The tension free energy is the line tension $I$ multiplied by the arc length of the disclination, which is the circular radius times the angle subtended.  The radius is $r=(y_\text{top}/2)+(w^2/(8y_\text{top}))$, and the angle is $\theta=4\arctan(2y_\text{top}/w)$.  Hence, the tension free energy is $F_\text{tension}=Ir\theta$, and its derivative with respect to $y_\text{top}$ is
\begin{equation}
\frac{\partial F_\text{tension}}{\partial y_\text{top}}=
\frac{I[2wy_\text{top}-(w^2-4y_\text{top}^2)\arctan(2y_\text{top}/w)]}{2y_\text{top}^2}.
\label{eq:Ftensionderiv}
\end{equation}

The twist free energy is associated with twist of the director between the bottom and top substrates, across the cell thickness $d$.  It is difficult to calculate the total twist free energy, integrated over the entire liquid crystal, but easier to calculate its derivative with respect to $y_\text{top}$.  If the center of the circle moves upward from $y_\text{top}$ to $y_\text{top}+dy_\text{top}$, then a narrow arc is transformed from Region~II to Region~I.  Remarkably, the anchoring orientation $\phi_B$ is uniform along this arc, with the value $\phi_B=\arctan(2y_\text{top}/w)$.  In Region~II, the twist angle is $(\phi_T-\phi_B)$, and hence the twist free energy density per area is $(K_{22}/(2d))(\phi_T-\phi_B)^2$.  In Region~I, the twist angle is reduced to $(\phi_T-\phi_B-\pi)$, and hence the twist free energy density per area is reduced to $(K_{22}/(2d))(\phi_T-\phi_B-\pi)^2$.  Under this change in $y_\text{top}$, the area transformed from Region~II to Region~I is
\begin{align}
\frac{\partial A}{\partial y_\text{top}}&=
\frac{1}{16y_\text{top}^3}[2wy_\text{top}(w^2+4y_\text{top}^2)\\
&\qquad\qquad-(w^4-16y_\text{top}^4)\arctan(2y_\text{top}/w)].\nonumber
\end{align}
Combining these factors, the derivative of twist free energy with respect to $y_\text{top}$ is
\begin{align}
\label{eq:Ftwistderiv}
\frac{\partial F_\text{twist}}{\partial y_\text{top}}&=
\frac{K_{22}\pi}{32dy_\text{top}^3}[\pi-2\phi_T+2\arctan(2y_\text{top}/w)]\\
&\qquad\times[2wy_\text{top}(w^2+4y_\text{top}^2)\nonumber\\
&\qquad\qquad-(w^4-16y_\text{top}^4)\arctan(2y_\text{top}/w)].\nonumber
\end{align}

We combine Eqs.~(\ref{eq:Ftensionderiv}) and~(\ref{eq:Ftwistderiv}) to obtain the derivative of total free energy with respect to $y_\text{top}$, and then set it equal to zero to find the minimum, giving the result in Eq.~(\ref{eq:DeltaPhiAndycBar}).  Formally, that equation gives $\delta\phi$ as a function of $y_\text{top}$.  However, it can be inverted either numerically or graphically to find $y_\text{top}$ as a function of $\delta\phi$.

\subsection{Peach-Koehler force}

To estimate the velocity of an expanding disclination loop, we must take the ratio of the total force (per length) to the drag coefficient (per length).

For a large disclination loop, we have $\phi_B\to\pi/2$, and hence the Peach-Koehler force has magnitude $(\pi K/d)(\delta\phi-\pi/2)$ in the outward direction.  In the same limit of large radius, the force of tension goes to zero.  Hence, the total force has magnitude $(\pi K/d)(\delta\phi-\pi/2)$.

If we neglect fluid flow, the drag coefficient~\cite{Imura1973,Tang2019} is given by $\zeta=\pi\gamma k^2 \log(r_\text{max}/r_\text{core})$.  Here, $\gamma$ is the rotational viscosity (in the director representation), $k=1/2$ is the topological charge of the disclination, $r_\text{core}$ is the disclination core radius, and $r_\text{max}$ is the distance from the disclination to a boundary.

Combining these expressions, we expect the velocity
\begin{equation}
v=\frac{f_\text{total}}{\zeta}=\frac{4K(\delta\phi-\pi/2)}{\gamma d\log(r_\text{max}/r_\text{core})}.
\end{equation}

In our nematic-order-tensor simulations, the cell thickness is $d=10$ (arbitrary units), and the single Frank elastic constant is $K=\frac{9}{2}S^2 L=3.0$.  The distance from the boundary is approximately $r_\text{max}=5$, and the core radius is approximately $r_\text{core}=1.6$.  The rotational viscosity $\gamma$ in the director representation is related to the corresponding viscosity $\Gamma$ in the tensor representation by $\gamma|\dot{\bm{n}}|^2=\Gamma\Tr\dot{\bm{Q}}^2$, and hence $\gamma=\frac{9}{2}S^2 \Gamma=1.3$.  Hence, our prediction for the disclination velocity becomes $v=0.81(\delta\phi-\pi/2)$.

\section{Competing interests}
The authors declare no competing interests.
\section{Acknowledgements}
Work supported by NSF CMMI-1663041.

%
\end{document}